\documentclass[aip,rsi,reprint]{revtex4-1}
\usepackage{graphicx}
\usepackage{amsmath}
\newcommand{\degrees}{$^\circ$}
\newcommand{\micron}{$\mu$m\ }
\begin{document}

\preprint{}

\title{Magneto-Optical Spectrum Analyzer}



\author{M. Helsen}
\email{Mathias.Helsen@UGent.be}
\affiliation{Department Solid State Sciences, Ghent University, Krijgslaan 281/S1, 9000 Ghent, Belgium}
\author{A. Gangwar}
\affiliation{Department of Physics, Universit\"{a}t Regensburg, Universit\"atsstrasse 31, 93040 Regensburg, Germany}
\author{A. Vansteenkiste} 
\affiliation{Department Solid State Sciences, Ghent University, Krijgslaan 281/S1, 9000 Ghent, Belgium}
\author{B. Van Waeyenberge}
\affiliation{Department Solid State Sciences, Ghent University, Krijgslaan 281/S1, 9000 Ghent, Belgium}


\date{\today}

\begin{abstract}
We present a method for the investigation of gigahertz magnetization dynamics of single magnetic nano elements.
By combining a frequency domain approach with a micro focus Kerr effect detection, 
a high sensitivity to magnetization dynamics with submicron spatial resolution is achieved. It allows 
 spectra of single nanostructures to be recorded. 
Results on the uniform precession in soft magnetic platelets are presented.
\end{abstract}


\maketitle 

\section{Introduction}
Magnetism at the sub-micrometer and nano scale attracts a great deal of interest
for both fundamental reasons and for their prospective use in logic and memory
applications. As not only the static properties, such as the magnetic magnetic domain
configuration, but also the dynamics properties on the sub-nanosecond timescale
(e.g. the resonances and magnetization switching) are strongly determined by the reduced
dimensionality, appropriate characterization techniques are required.   
The conventional method for high frequency characterisation of magnetic
systems is the cavity based ferromagnetic resonace technique. However, nanostuctures
can not be studied in remanence as the fixed frequency operation requires
the bias field to be swept.  To address this problem, different techniques
have been developed. Dependent on how the magnetic response is detected they
can be divided in two categories.   Vector Network Analyzer Ferromagnetic Resonance
(VNA-FMR \cite{neudecker06,kalarickal06}) and Pulsed Inductive Microwave Magnetometry
(PIMM \cite{kos02,silva99}) detect the resonance electrically and in Time Resolved
Magneto-Optical Kerr Effect (TR-MOKE \cite{acremann01}) and Time Resolved Scanning
Transmission X-Ray Microscopy experiments (STXM \cite{bolte08,vansteenkiste09,kammerer11}) optical
detection is used. The optical methods have a high sensisitivity to
detect the signal of a single magnetic microstructure, but have a much more complicated set-up than the electrical methods,
and require a femtosecond laser or pulsed X-ray source.  On the other hand, detection in the
frequency domain, like conventional FMR and VNA-FMR can achieve much higher signal-to-noise ratios \cite{neudecker06}. 
Here we present an approach which combines the frequency domain method with optical detection:
the Magneto-Optical Spectrum Analyzer (MOSA).

This method is a hybrid method between VNA-FMR and TR-MOKE in the context of measurement abilities and construction. 
TR-MOKE makes use of a pulsed laser to probe the magnetization ($\vec{M}$)
through the magneto-optical Kerr effect\cite{argyres55,zak90,polisetty08,zvezdin97,qiu00} at regular intervals, while the sample 
is excited using e.g. another laser pulse or a microwave signal generator. By shifting the arrival time of
the probe pulses with respect to the excitation, time domain information can be gathered. 
This method, on the other hand
probes the magnetization using a continuous wave laser and measures (again through the Kerr effect) the fast change
of magnetization with an ultrafast photodiode in the frequency domain. 
\section{Description of the set-up}
\begin{figure}
    \includegraphics[width=8.5cm]{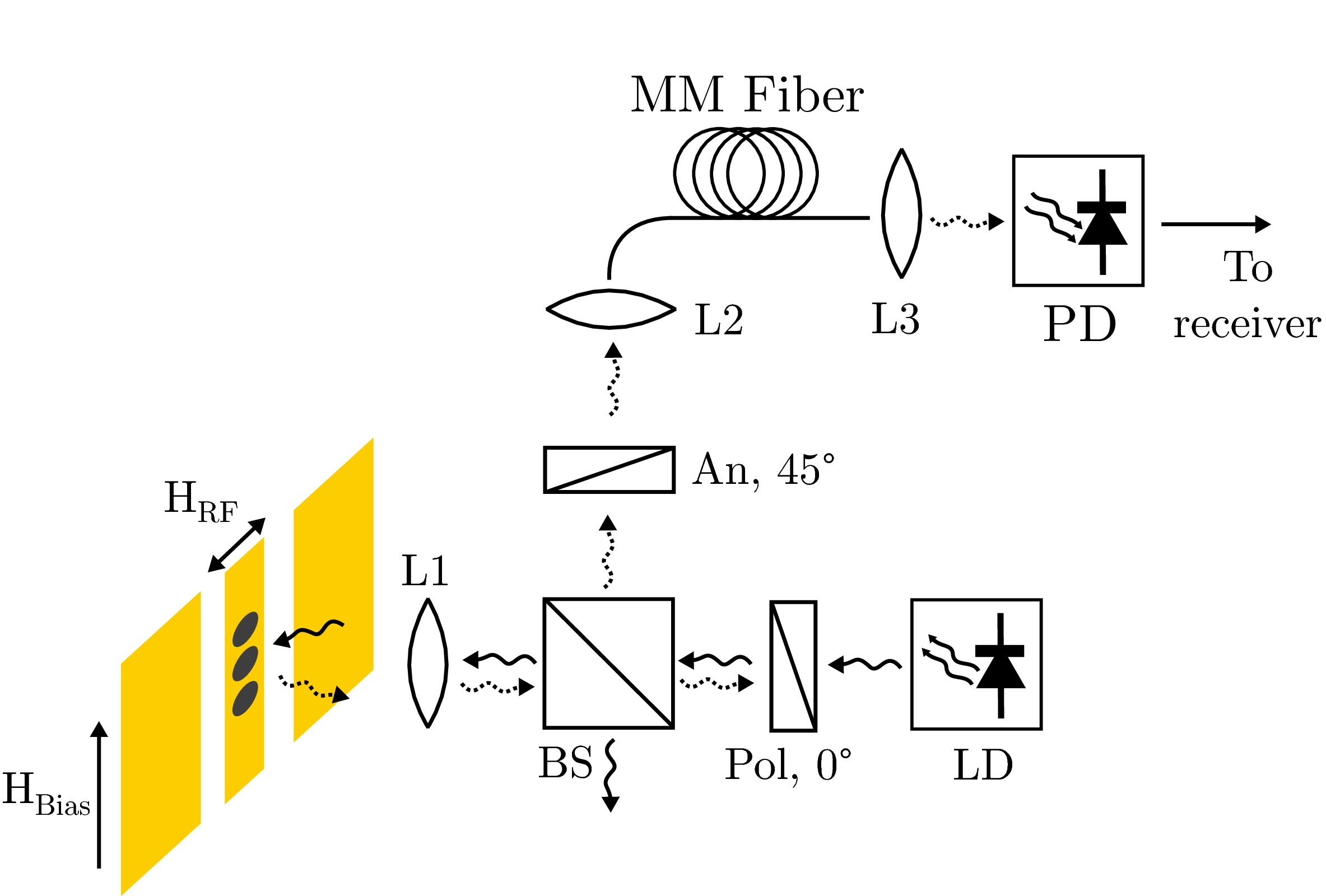}
    \caption{A basic sketch of the optical part of the set-up, used for polar Kerr detection. LD is a laser diode, Pol a polarizer, BS a non-polarizing beamsplitter,
    L1 an objective lens, An an analyzer, L2 and L3 aspheric lenses and PD an ultrafast photodiode. The angle between the transmission axis of Pol and An is 45\degrees.}
    \label{setup}
\end{figure}

Shown in Fig.\ref{setup} is the optical layout of our set-up. Light (660nm) from a laser diode (LD) is linearly
polarized using a polarizer (Pol) and passes through a non-polarizing beamsplitter (BS). An objective lens (L1)
focusses it to a diffraction limited spot ($\approx 500$nm) on the sample, where microwave interconnects provide RF current for the excitation 
and an electromagnet can provide a bias field of up to 50mT.

The reflected light is collected by the same objective lens and redirected by the beamsplitter onto an analyzer (An).
The tranmission axis of this analyzer is set at an angle of 45\degrees\  with respect to the transmission axis of the first
polarizer. The beam is then focussed using an aspheric lens (L2) onto a multimode optical fiber (MM Fiber), 
transporting it over a large distance (50 m in our case) to an ultrafast photodiode (PD) with a bandwidth of approximately 12GHz. At this point intensity variations are
measured.

The objective lens is mounted on a piezo stage to allow scanning of the probe beam over the sample surface. By recording
the DC reflected light intensity, we can image the sample and correctly focus and position the probe beam on the sample.
For this purpose the light is deflected with a mirror towards a conventional photodiode (both not shown for clarity).

The out-of-plane magnetization dynamics are measured via the polar Kerr effect. When reflecting off the sample, linearly polarized lights
gains both an ellipticity ($\epsilon_\text{K}$) and a rotation
of the major polarization axis, known as the Kerr angle ($\theta_\text{K}$). For out-of-plane saturated Permalloy, the polar Kerr angle is 
typically 1 mrad\cite{veis12}. Both the sign and magnitude of this angle depend on the sign and magnitude of 
the out-of-plane magnetization of the probed area.

The reflected light is analyzed with the polarizer at an angle of 45\degrees, thus yielding an intensity of $I = I_0(1/2+\theta_\text{K})$, where
$I_0$ is the intensity before the analyzer. Placing the analyzer at 45\degrees\  maximizes the signal.

\begin{figure}
    \includegraphics[width=8.5cm]{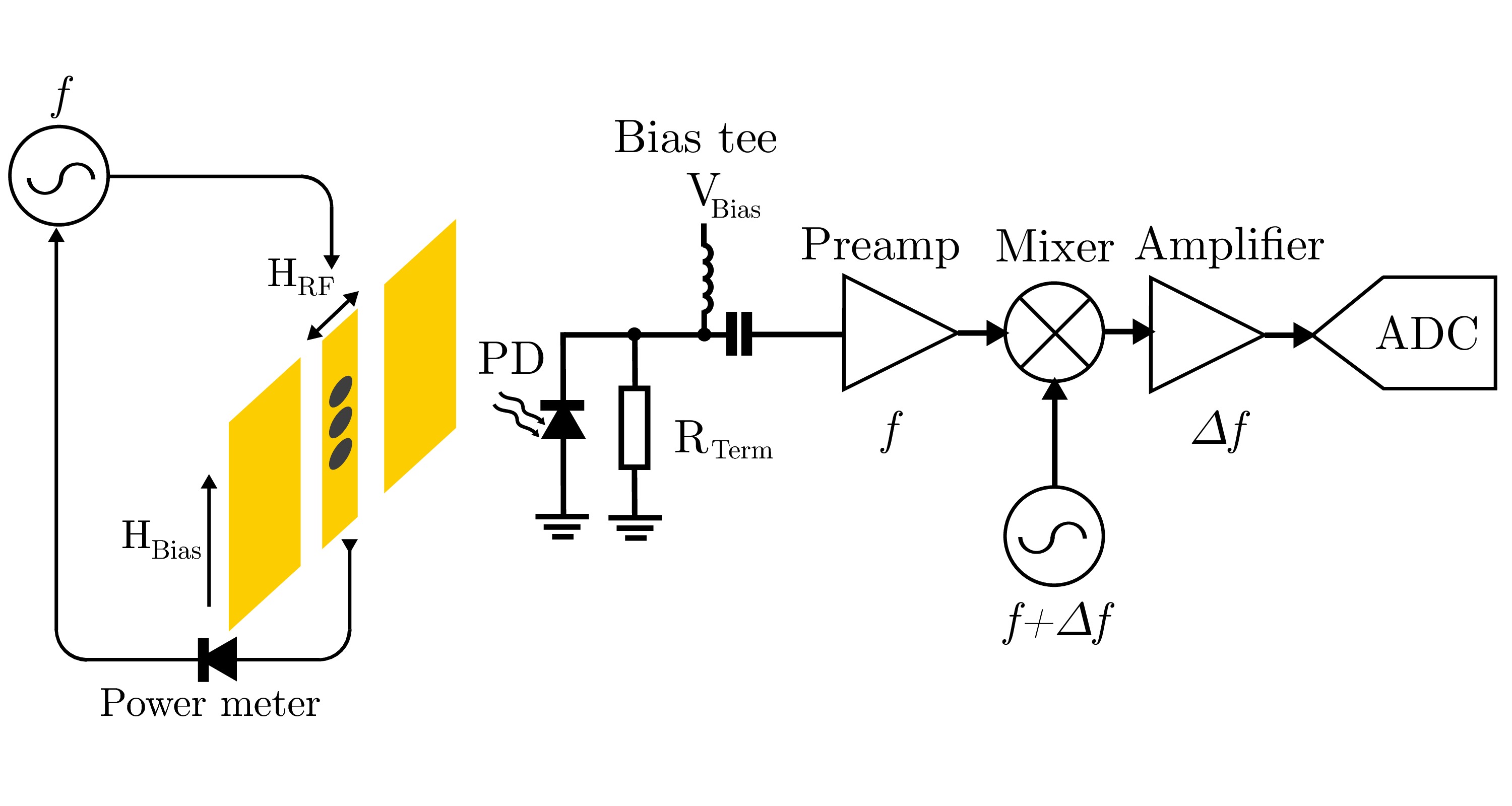}
    \caption{A simplified schematic of the electrical part of the set-up. A diode detector connected to the sample helps in keeping the power transmitted through the sample level when the frequency is varied. Also shown are the photodiode (PD), termination resistor ($R_\text{Term}$), bias tee, RF preamplifier, frequency mixer, low frequency amplifier and ADC. Both signal generators produce a tone in the microwave range ($f$), but are slightly offset by frequency $\Delta f$ in the kHz range.
    \label{setup2}}
\end{figure}

The electrical part of the set-up is shown in Fig. \ref{setup2}.
One signal generator produces the RF current at frequency $f$ used for
exciting the sample. The RF power through the sample is kept level by using a diode detector measuring the power coming out
of the sample. The bias tee provides 
the necessary bias voltage and shunts the DC photocurrent. The DC photocurrent flowing out of the bias tee is measured for monitoring
purposes.

Assuming a linear dependence of the light intensity on the magnetization \cite{argyres55,zvezdin97}, we can estimate the AC current induced in the photodiode
due to magnetization dynamics as $\delta I \approx I_\text{DC} \theta_\text{K,max} \delta m_z$, where $\theta_\text{K,max}$ 
is the Kerr angle at saturation ($M_z = M_\text{S}$),
$\delta m$ the reduced out-of-plane magnetization ($M_z/M_\text{S}$) and $I_\text{DC}$ the DC photocurrent.

The AC photocurrent is passed on to the
low noise preamplifier, which increases the signal level by 30dB.
After this preamplifier, a mixer downconverts this high frequency signal to a frequency in the audio range. To this end
a second signal generator, phase locked to the first one, produces a high frequency signal
at a frequency offseted by several kilohertz ($\Delta f$) with respect to the excitation frequency $f$. Thus, the signal that the mixer produces is
at the frequency $\Delta f$. It is further amplified by a second amplifier and finally sampled using a high end ADC. 
A computer records the incoming data from the ADC and performs an FFT to compute the signal strength at $\Delta f$.

\section{Comparison with other methods}
To estimate the detection limit of the set-up, we compare the typical signal levels to the fundamental noise present. There are
three main contributions to this noise\cite{horowitz89}:
\begin{itemize}
    \item a contribution from the DC photocurrent, known as shot noise, with a current noise density given by $i_\text{noise} = \sqrt{2 q I_\text{DC}}$;
    \item a contribution from the noise intrinsic to the diode, quantified by the Noise Equivalent Power (NEP);
    \item and a contribution from the terminating resistor, $R_\text{Term}$ (50$\Omega$), known as Johnson noise, with a voltage noise density given by $v_\text{noise} = \sqrt{4 kTR_\text{Term}}$.
\end{itemize} We assume that these are sources of white noise and that all other possible sources generate much less
noise in the frequency range of interest.

If we compare this noise with the signal strength, we arrive at the following expression for the Signal-to-Noise Ratio (SNR) at the input of the preamp:
\begin{equation}
    \text{SNR} = 10 \text{log}_{10} \frac{ I_\text{DC}^2 \theta_\text{K}^2 \langle m_z^2 \rangle }{B (2q I_\text{DC} + 4kT/R_\text{Term} + (\mathcal{R}\cdot \text{NEP})^2)},
\end{equation} where $B$ is the measurement bandwidth and $\mathcal{R}$ the responsivity of the photodiode (A/W (at a specific wavelength)).
Typical values for our set-up would involve $I_\text{DC} = 100.0\mu\text{A}$, $B = 1\text{Hz}$ and $\langle m_z^2 \rangle = 10^{-4}$.
The Johnson noise is the largest contribution ($1.8 \cdot 10^{-11}$ A/$\sqrt{\text{Hz}}$), followed by the shot noise ($5.7 \cdot 10^{-12}$ A/$\sqrt{\text{Hz}}$). 
In comparision, the NEP for the photodiode we have used is negligible (only $4.5 \cdot 10^{-17}$  A/$\sqrt{\text{Hz}}$). Because the contribution from shot noise
is still smaller than the Johnson noise, the SNR can be significantly improved by increasing the light intensity incident on the photodiode. Only when the photocurrent
reaches 1 mA (equivalent to 10 mW optical power on the diode) does the shot noise exceed the Johnson noise. However increasing light intensity
also entails heating up to sample, even to above the Curie temperature.
A further increase in SNR
could be obtained by cooling the detector, lowering the Johnson noise.

Adding these terms we find a SNR
of approximatly 34dB. In addition, the first amplifier adds a noise figure of 2dB, thus the final SNR is about 32dB. 
But as the signal itself is on the order of -130dBm or $5\cdot 10^{-14}$ mW, absolute care in handling the microwave signals is still required. 

\begin{figure}
    \includegraphics[width=8.5cm]{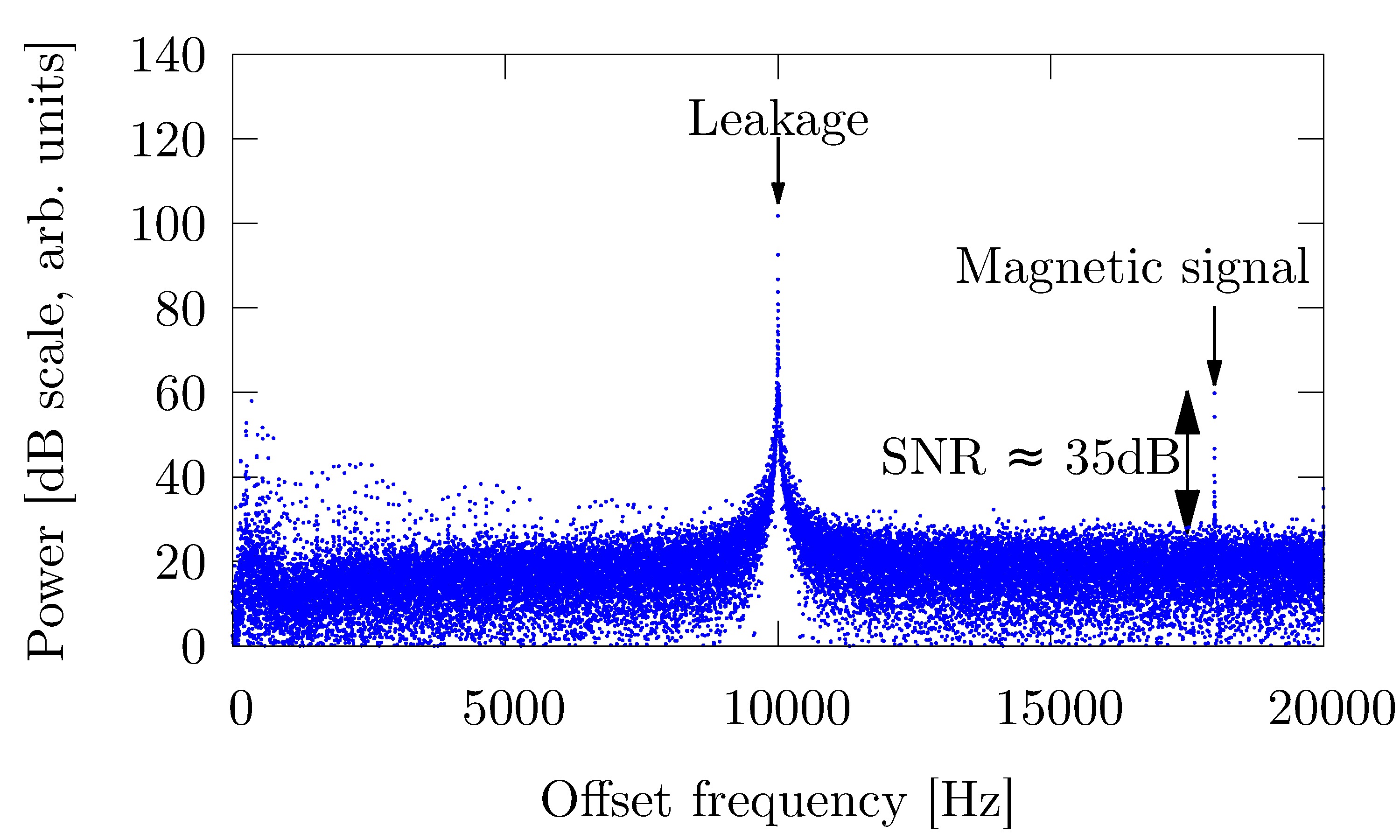}
    \caption{Shown is a FFT when both a magnetic signal and a direct coupling are present. In this particular case
    a Permalloy disc with a diameter of 20\micron was excited at 4GHz with +10dBm of power and biased with a field of 20mT. The leakage was caused by 
    a microstrip, which was properly terminated and carried the same amount of power as used for the excitation, but at a slightly different frequency.\label{leakage}}
\end{figure}

To illustrate this, we have excited a sample with enough RF power (+10dBm) so that the it would precess with 
$\delta m_z \approx 0.01$. The same amount of RF power but at a slightly offseted frequency
 was sent through a microstrip next to the receiver, to allow for a comparison between the magnetic signal and the direct coupling
from the excitation to the receiver. The result in Fig. \ref{leakage} shows that direct coupling is much stronger than the magnetic signal and that
our estimate of the SNR is quite accurate. To this end, the detection
has been separated by a large distance from the excitation. 
Low frequency $1/f$ noise is not showing due to a low frequency cut-off characteristic of the low frequency amplifier,
but the noise floor increases below 5kHz.

Our method is complementary to both VNA-FMR and TR-MOKE. To the former, we add the advantage of spatial selectivity. This means
that several magnetic structures can be fabricated in each others vicinity and we are still able to probe each one separately, or that the
spatial variation of the magnetization dynamics can be analyzed, in contrast to VNA-FMR.

Where TR-MOKE is a time domain method, we are measuring directly in frequency domain, thus it is easier to measure resonance curves
directly and not have to rely on Fourier transforms of time domain data. 
In comparison to pulsed methods, we have independent control of frequency and amplitude, resulting in non-ambiguous spectra.
Further, our method allows us to probe the signal at any arbitrary frequency,
something that is difficult to do with stroboscopic time domain measurement methods.

One last advantage of our method is that it does not require a femtosecond pulsed laser, high end oscilloscope or vector network analyzer, thus eliminating a large cost.
\section{Experimental details}
As an illustration we have measured the uniform precession of magnetization in Permalloy discs with a 20\micron diameter. 
The samples were fabricated on a silicon substrate. Structure definitions were made with electron beam lithography and the lift-off technique.
After the last metal deposition an ALD coating of Al$_2$O$_3$ was deposited. The samples were then wire bonded to a high frequency substrate.
An example of such a sample is shown in Fig. \ref{figSample}.
\begin{figure}
    \includegraphics[width=6cm]{fmr_sample.jpg}
    \caption{A SEM micrograph of a sample used for measurements. The dots are 75\micron and 20\micron in diameter. \label{figSample}}
\end{figure}

\begin{figure}
    \includegraphics[width=8.5cm]{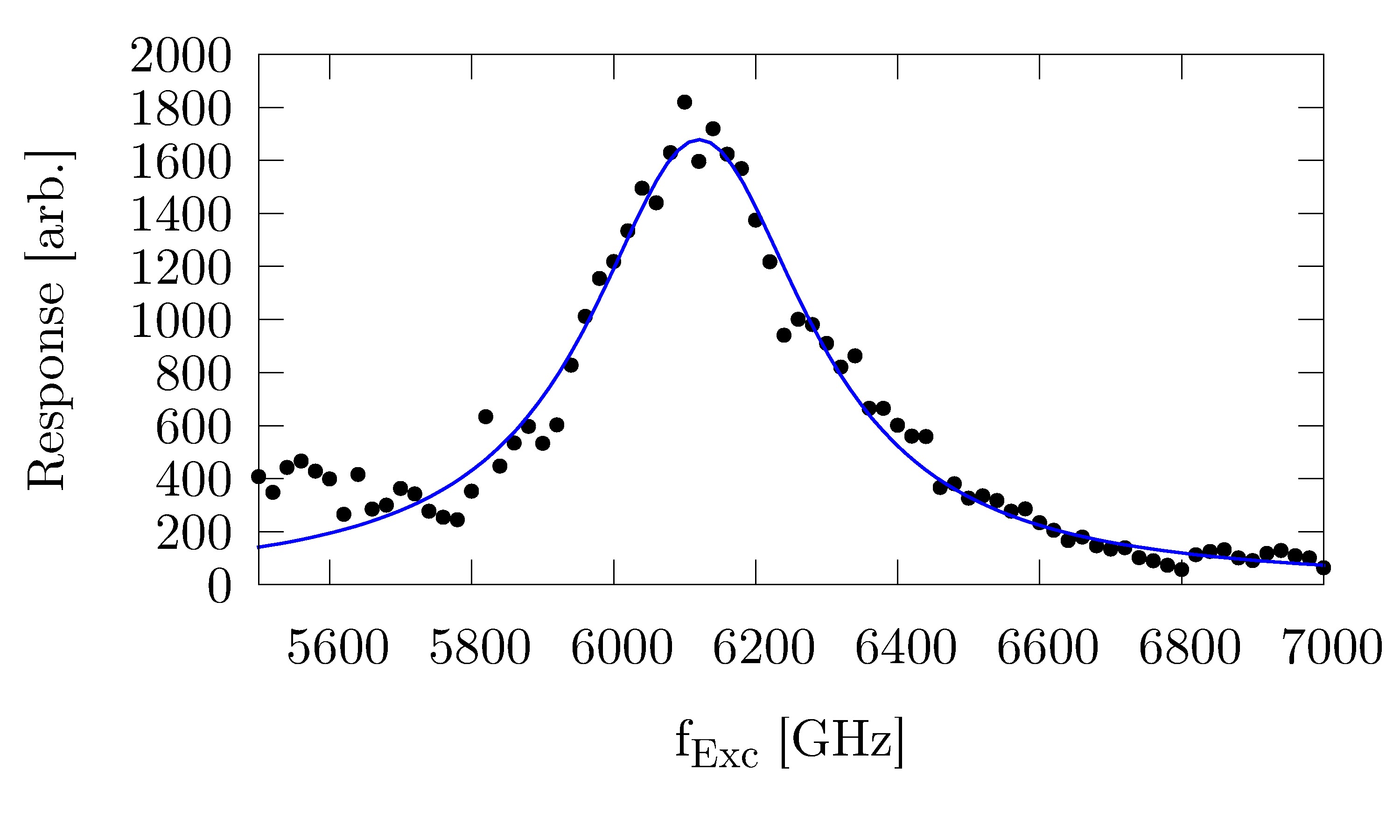}
    \caption{The resonance curve for a 20\micron diameter, 50nm thickness Permalloy disc in a field of 45mT. The solid
    line is the Lorentzian fit.\label{resonance}}
\end{figure}

An example of a resonance curve
where the frequency is swept at a fixed field of 45mT, is shown in Fig. \ref{resonance}. The peak was fitted
to a Lorentz curve and the resulting resonance frequency was determined to be 6120$\pm$2 MHz, with a linewidth of 188$\pm$4 MHz. 
This illustrates that the linewidth can be accuratly measured on single microscopic elements.

\begin{figure}
    \includegraphics[width=8.5cm]{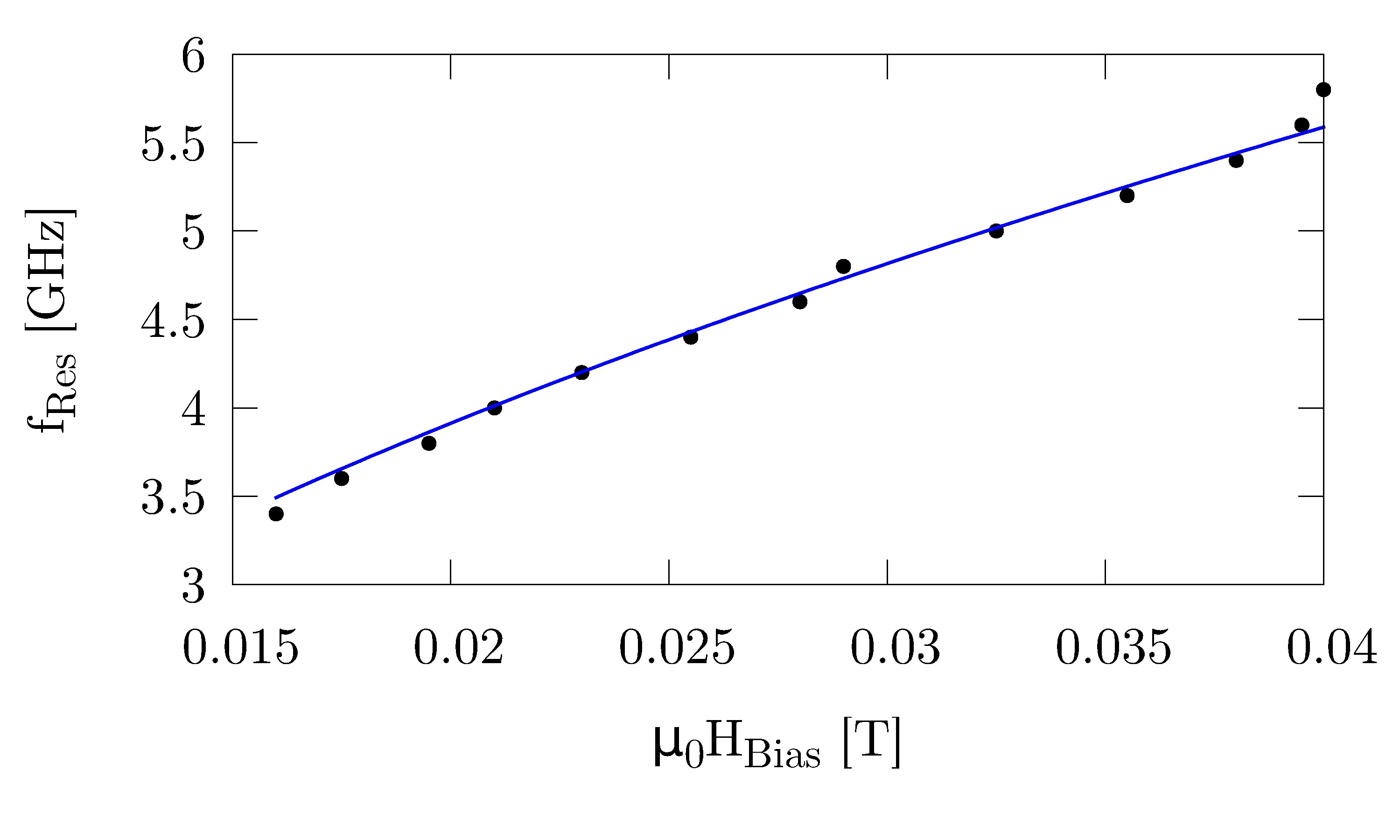}
    \caption{The evolution of the resonance peak of the uniform precession when the field is increased. 
	The Kittel equation for a thin film is shown as a black line. \label{kittel}}
\end{figure}

The detector itself has a frequency dependence, making the recorded spectrum a product of the magnetic response and the
detector response. To investigate this effect, a frequency sweep was performed for a different number of bias fields and the resonance frequency was determined for each.
In Fig. \ref{kittel} the resulting datapoints are compared with the Kittel equation, which for an infinitly thin film is given by
$f_\text{Res} = 28.0 \sqrt{ \mu_0 H_\text{Bias} (\mu_0 H_\text{Bias} + \mu_0 M_s) }$ GHz/T.
 Literature values have been used
for calculating the Kittel equation ($\mu_0 M_s = 1.04$T)\cite{coey10}. The datapoints are in good agreement with the Kittel equation, 
ruling out strong frequency variations in the detection which might interfere with measurements.

\begin{figure}
    \includegraphics[width=8.5cm]{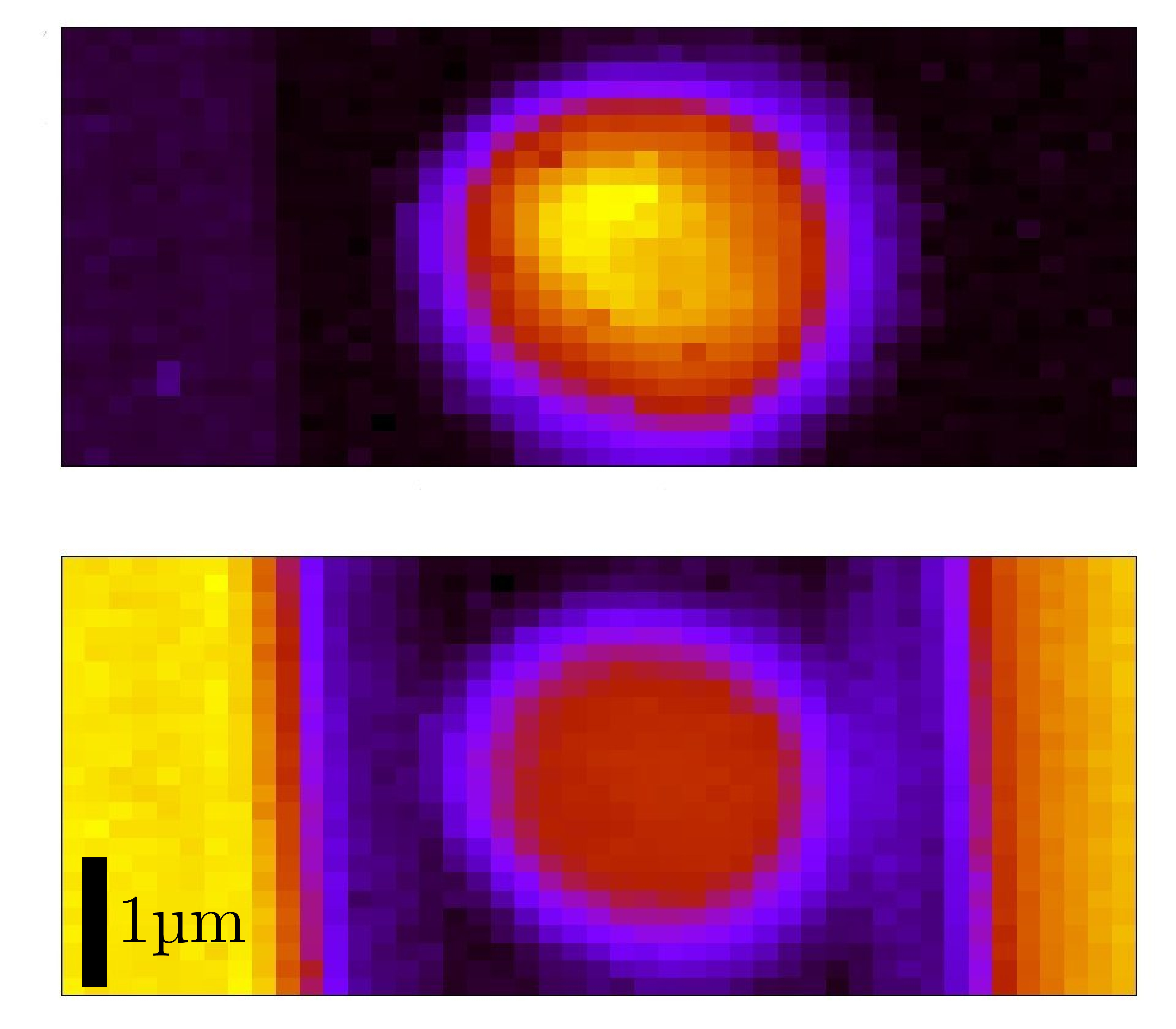}
    \caption{At the top an image generated using the magnetic response of a uniform resonance in a 3\micron dot at 6GHz. Shown
    at the bottom is an image generated using reflectivity data that was collected simultaneously.
    \label{sample2}}
\end{figure}

Finally, the magnetic signal can also be used for imaging purposes; when the laser beam is scanned over the sample using the piezo stage
and the magnetic signal at each point is recorded. This can be compared with the reflectivity,
which is acquired simultaneously. When the sample is excited at resonance, the contrast is highest. An example
is shown in Fig. \ref{sample2}. Here we compare the reflectivity (clearly showing the Au CPW, Si substrate and Permalloy discs) with
the magnetic signal showing only the Permalloy disc.

\section{Conclusion}
We have developed a method of probing magnetization dynamics at the multiple-GHz range using 
a frequency domain method that offers spatial sensitivity. The set-up is relatively simple, yet allows
for high quality measurements, thus enabling a fast exploration of excitation parameters. We have illustrated
that our method can yield quantitative results using uniform excitation on single microscopic elements.
\begin{acknowledgments}
Mathias Helsen, Arne Vansteenkiste and Bartel Van Waeyenberge acknowledge funding by FWO-Vlaanderen and BOF-UGent.
\end{acknowledgments}

\bibliographystyle{ieeetr}
\bibliography{biblio}

\end{document}